%% file: main.tex
\documentclass[sigconf,natbib=true]{acmart}
\usepackage{multirow}
\usepackage{float}
\usepackage{url}
\usepackage{algorithm}
\usepackage{algpseudocode}
\usepackage{bm}
\usepackage{graphicx}
\usepackage{caption}
\usepackage[skip=0cm,list=true,labelfont=it]{subcaption}
\usepackage{enumitem}

\setlist[itemize]{leftmargin=8pt,labelindent=0pt}

\newtheorem{assumption}{Assumption} 
\usepackage{amsthm}
\usepackage{bbm}
\newcommand{\U}{\mathcal{U}}
\newcommand{\I}{\mathcal{I}}
\newcommand{\D}{\mathcal{D}}

\newcommand{\E}{\mathbb{E}}
\newcommand{\TV}{D_{\mathrm{TV}}}
\newcommand{\Hb}{h_{\mathrm b}}

\theoremstyle{definition}
\newtheorem{definition}{Definition}[subsection]

\AtBeginDocument{%
  \providecommand\BibTeX{{%
    \normalfont B\kern-0.5em{\scshape i\kern-0.25em b}\kern-0.8em\TeX}}}

\copyrightyear{2027}
\acmYear{2027}
\setcopyright{cc}
\setcctype{by}
\acmConference[WSDM '27]{Proceedings of the Twentieth ACM International Conference on Web Search and Data Mining}{February 15--19, 2027}{Hong Kong, Hong Kong}
\acmBooktitle{Proceedings of the Twentieth ACM International Conference on Web Search and Data Mining (WSDM '27), February 15--19, 2027, Hong Kong, Hong Kong}
\acmDOI{10.1145/3862322.3863878}
\acmISBN{979-8-4007-3117-4/2027/02}

\begin{document}

\title[How User Overlap Shapes Diversity in Cross-Domain Recommendation]{Beyond Knowledge Transfer: How User Overlap Shapes Diversity in Cross-Domain Recommendation}

\author{Yuhan Zhao}
\affiliation{
  \institution{Hong Kong Baptist University}
  \city{Hong Kong}
  \country{China}
}
\email{csyhzhao@comp.hkbu.edu.hk}

\author{Weixin Chen}
\affiliation{
  \institution{Hong Kong Baptist University}
  \city{Hong Kong}
  \country{China}
}
\email{cswxchen@comp.hkbu.edu.hk}

\author{Shu Chen}
\affiliation{
  \institution{Shenzhen University}
  \city{Shenzhen}
  \country{China}
}
\email{chenshu20@email.szu.edu.cn}

\author{Li Chen}
\affiliation{
  \institution{Hong Kong Baptist University}
  \city{Hong Kong}
  \country{China}
}
\email{lichen@comp.hkbu.edu.hk}

\input{0.abstract}

\ccsdesc[500]{Information systems}
\ccsdesc[500]{Information systems~Recommender systems}

\keywords{Cross-domain recommendation, non-overlapping users, diversity}
\maketitle

\input{1.introduction}

\input{3.preliminary}
\input{4.methodology}

\input{5.experiment}

\input{2.related_work}
\input{7.conclusion}
\input{acks}

\clearpage
\input{6.ethic}

\balance
\bibliographystyle{ACM-Reference-Format}
\bibliography{paper}

\clearpage

\end{document}

%% file: 0.abstract.tex
\begin{abstract}
A common paradigm in cross-domain recommendation (CDR) transfers knowledge through users shared by source and target domains. Although this overlap bridge is effective in improving accuracy, its effect on diversity remains unclear. We identify two bottlenecks: the bridge can be too narrow when overlapping users cover limited target-domain tastes, and too loud when shared signals dominate the sparse personal evidence of non-overlapping users. To address both, we propose DivCDR, a model-agnostic Push--Pull framework. Pull constructs confidence-gated virtual positives for overlapping users and optimizes their exposure-entropy gain, broadening the preferences carried by the bridge. Push uses overlap-density-guided negative sampling to move non-overlapping users away from overlap-dense regions. Experiments on three datasets and five representative CDR backbones show that DivCDR improves recommendation diversity while preserving or improving accuracy in most settings. 
The implementation is publicly available at \url{https://github.com/weixinchen98/DivCDR}.
\end{abstract}

%% file: 1.introduction.tex
\section{Introduction}
\label{sec:introduction}

Cross-domain recommendation (CDR) improves recommendation quality in a target domain by leveraging auxiliary knowledge from one or more source domains~\cite{LCL24,XLW22,hu2026ids}. A widely adopted strategy relies on overlapping users, users with interactions in both the source and target domains, as bridges for knowledge transfer~\cite{WYZ25,chen2025lasso}.

By transferring source-domain knowledge, CDR has achieved substantial improvements in recommendation accuracy~\cite{hu2026ids}. As the field matures, however, an important question arises~\cite{khan2017cross,cantador2015cross,fernandez2012cross}: \textbf{beyond accuracy, can cross-domain knowledge improve recommendation diversity?} Intuitively, incorporating auxiliary knowledge from another domain should broaden user exposure and thereby mitigate filter bubbles and echo chambers. Existing works~\cite{fernandez2016accuracy,chen2026divcdsr}, however, indicate a counterintuitive pattern: CDR often yields limited diversity gains and can even make recommendations more homogeneous. The mechanisms underlying this phenomenon remain insufficiently understood. To address this gap, we investigate the following question: {\textbf{What factors govern diversity in CDR?}} Through theoretical analysis and empirical studies (detailed in Section~\ref{sec:theory}), we characterize this phenomenon through two bottlenecks associated with the knowledge-transfer ``bridge'':

\begin{itemize}
    \item \textbf{Bottleneck I: The bridge is too narrow.}
    The first bottleneck concerns {what can be transferred}. Because overlapping users provide the identity-linked pathways between domains, the resulting transfer bridge can carry only the range of preferences represented by those users.

    \item \textbf{Bottleneck II: The bridge is too loud.}
    The second bottleneck concerns {what happens after transfer}. A non-overlapping user has no source-domain history. If that user also has few interactions in the target domain, the transferred bridge signal can overwhelm the limited evidence of their individual interests. Consequently, the user's learned representation may remain near the representations of overlapping users, making recommendations across non-overlapping users more homogeneous.
\end{itemize}

These bottlenecks are distinct but related. The former is a menu problem: does the bridge carry a sufficiently broad range of preferences? The latter is a weight problem: does the transferred signal dominate the personal evidence of non-overlapping users?

To address these bottlenecks, we propose \textbf{DivCDR}, a model-agnostic Push--Pull framework. DivCDR builds on the complementary geometric effects of positive and negative samples in recommendation learning: positive samples attract a user representation, whereas negative samples repel it~\cite{RFG12,HYC17,ZZY22}. Following this principle, DivCDR introduces two targeted interventions.
For \textit{overlapping users}, \textbf{Pull} broadens the range of preferences carried by the bridge. It perturbs an observed positive item to construct a user-conditioned virtual positive and directly optimizes a differentiable diversity-gain objective. 
For \textit{non-overlapping users}, \textbf{Push} weakens the dominance of the bridge. It estimates cocoon strength using an overlap-density surrogate over overlapping-user representations and samples a negative item whose induced repulsive update is aligned with the local escape direction. The resulting repulsive update is designed to move the user representation toward a lower-density, and hence less homogenized, region. Our main contributions are as follows:
\begin{itemize}
    \item We provide mathematical analysis and experiments for two overlap-conditioned diversity failure modes: concentrated exposure among overlapping users and overlap-dominated representations for non-overlapping users.
    \item We introduce DivCDR, a model-agnostic framework whose Pull and Push components directly address the two bottlenecks without modifying the underlying CDR architecture. Pull broadens the transfer menu through user-conditioned virtual positives, whereas Push weakens information cocoons through overlap-density-guided negative sampling.
    \item Across three real-world datasets and five representative CDR backbones, DivCDR improves recommendation diversity while preserving or improving accuracy in most cases.
\end{itemize}

%% file: 3.preliminary.tex
\section{Dual Diversity Bottlenecks in CDR}
\label{sec:theory}

\subsection{Setup and Diversity Quantities}
\label{subsec:setup}

\begin{definition}[Partially overlapping CDR]
A CDR problem consists of a source domain
$\D_s=(\U_s,\I_s,Y_s)$ and a target domain
$\D_t=(\U_t,\I_t,Y_t)$.
The two target-user groups are
\begin{equation}
    \U_o=\U_s\cap\U_t,
    \qquad
    \U_n=\U_t\setminus\U_o,
\end{equation}
where $\U_o$ and $\U_n$ denote overlapping and non-overlapping users, respectively. We consider the prevalent CDR setting in which only a subset of users appear in both domains, with the goal of improving target-domain ranking performance~\cite{ZWC21,ZLZY23}.
\end{definition}

Let $n=|\I_t|$. For a target-domain user $u$, let $L_u^K$ be the top-$K$ list.
For a group $G\subseteq\U_t$, its normalized exposure distribution is
\begin{equation}
    q_G(i)=\frac{1}{K|G|}\sum_{u\in G} \mathbbm{1} [i\in L_u^K],
    \qquad i\in\I_t.
    \label{eq:group_exposure}
\end{equation}
We write $q_o=q_{\U_o}$ for overlap exposure and measure its effective support by
\begin{equation}
    H(q_o)=-\sum_{i\in\I_t}q_o(i)\log q_o(i),
    \qquad
    N_{\mathrm{eff}}(q_o)=\exp(H(q_o)).
    \label{eq:effective_support}
\end{equation}
Here, $q_o(i)$ is the relative frequency with which item $i$ is recommended to overlapping users. Larger entropy $H(q_o)$ indicates a more dispersed exposure distribution, and $N_{\mathrm{eff}}(q_o)$ is the corresponding effective number of equally exposed items. Although recommendation diversity can be measured in several ways, this entropy-based formulation enables a tractable theoretical analysis.

\subsection{Bottleneck I: A Narrow Bridge}
\label{subsec:bridge_bottleneck}    
Overlapping users in $\mathcal{U}_o$ provide the identity-linked pathways between the source and target domains. We therefore use $q_o$ as an observable proxy for the preference menu represented by these pathways.
If most of its probability mass lies on a small set $S_o$, this observable menu remains narrow even when the target catalog is large. The following bound makes the diversity ceiling explicit.

\begin{proposition}[Bridge entropy ceiling]
\label{prop:bridge_entropy}
Let $S_o\subset\I_t$ have size $s$.
Assume $1\le s<n$.
If $\delta=q_o(S_o^c)$, then
\begin{equation}
    H(q_o)
    \le
    \Hb(\delta)
    +(1-\delta)\log s
    +\delta\log(n-s),
    \label{eq:bridge_entropy_bound}
\end{equation}
where $\Hb(\delta)=-\delta\log\delta-(1-\delta)\log(1-\delta)$, with the convention $0\log 0=0$.
Consequently, $N_{\mathrm{eff}}(q_o)$ is at most the exponential of the right-hand side.
\end{proposition}

\begin{proof}
Draw $I\sim q_o$ and define $Z=\mathbbm{1}[I\in S_o]$.
The chain rule gives
\[
H(I)=H(Z)+(1-\delta)H(I\mid Z=1)+\delta H(I\mid Z=0).
\]
The conditional entropies are maximized by uniform distributions on their respective supports, so they are at most $\log s$ and $\log(n-s)$.
Substitution yields~\eqref{eq:bridge_entropy_bound}.
\end{proof}

\paragraph{Consequence.}
When most overlap exposure lies in a small $\mathcal S_o$, the observable overlap-exposure proxy $q_o$ has low effective support, even if the target catalog is large. This result formalizes the notion that the overlap-supported exposure is ``too narrow.'' It characterizes the concentration of an observable bridge-supported component, rather than a universal menu containing all information
that can be transferred by every CDR backbone. Under models in which the transferred component is substantially shaped by $q_o$, such concentration can constrain the diversity inherited by non-overlapping users and therefore motivates increasing diversity directly for $\mathcal U_o$.

\subsection{Bottleneck II: An Overpowering Bridge}
\label{subsec:cocoon_bottleneck}

We next ask when the bridge becomes too loud for $\U_n$.
A non-overlapping user must combine a shared, overlap-induced belief with evidence from the user's own target-domain history.
When that history is short, it may not supply enough evidence to move outside the region preferred by overlapping users.
The Bayesian formulation below provides an analytical lens on this learned inductive bias; DivCDR itself does not require a Bayesian backbone.

\paragraph{Preference and overlap-induced belief.}
Let $(\Theta,\mathcal B)$ be a measurable target-domain preference space.
A non-overlapping user has a latent preference $\theta^\star\in\Theta$ and provides
$m$ target-domain interactions $Y^{(m)}$ with likelihood
$L_m(\theta)=P_\theta^{(m)}(Y^{(m)})$.
Shared cross-domain training induces a belief $\Pi_O$ whose high-mass region
$N_o\subseteq\Theta$ is supported by overlapping users.
After observing $Y^{(m)}$, the induced posterior is
\begin{equation}
\Pi_m(B\mid Y^{(m)};\Pi_O)
=
\frac{\int_B L_m(\theta)\Pi_O(d\theta)}
{\int_\Theta L_m(\theta)\Pi_O(d\theta)}.
\label{eq:posterior}
\end{equation}

\begin{definition}[Finite-sample CDR cocoon]
\label{def:cocoon}
A non-overlapping user is $(N_o,\varepsilon)$-cocooned when
\begin{equation}
    \Pi_m(N_o\mid Y^{(m)};\Pi_O)\ge 1-\varepsilon.
    \label{eq:cocoon_definition}
\end{equation}
Thus, most of the inferred preference mass remains in the overlap-supported region after incorporating the user's target-domain evidence.
\end{definition}

\begin{assumption}[Finite-sample evidence budget]
\label{ass:evidence}
For the realized $Y^{(m)}$, suppose $\inf_{\vartheta\in N_o}L_m(\vartheta)>0$ and
\begin{equation}
    \frac{\sup_{\theta\in N_o^c}L_m(\theta)}
         {\inf_{\vartheta\in N_o}L_m(\vartheta)}
    \le \exp(m\kappa)
    \label{eq:evidence_budget}
\end{equation}
for some $\kappa\ge0$.
This condition bounds how much $m$ target-domain observations can change the odds between the two regions.
\end{assumption}

\begin{theorem}[Overlap concentration versus target-domain evidence]
\label{thm:cocoon}
If $\Pi_O(N_o^c)\le\delta<1$ and Assumption~\ref{ass:evidence} holds, then
\begin{equation}
    \Pi_m(N_o^c\mid Y^{(m)};\Pi_O)
    \le
    \frac{r_m}{1+r_m},
    \qquad
    r_m=\frac{\delta}{1-\delta}\exp(m\kappa).
    \label{eq:cocoon_bound}
\end{equation}
In particular, the user is $(N_o,\varepsilon_m)$-cocooned for
$\varepsilon_m=r_m/(1+r_m)$.
\end{theorem}

\begin{proof}
Posterior odds can be bounded as
\begin{align}
\frac{\Pi_m(N_o^c\mid Y^{(m)};\Pi_O)}
     {\Pi_m(N_o\mid Y^{(m)};\Pi_O)}
&=
\frac{\int_{N_o^c}L_m(\theta)\Pi_O(d\theta)}
     {\int_{N_o}L_m(\vartheta)\Pi_O(d\vartheta)}
\\
&\le
\frac{\sup_{\theta\in N_o^c}L_m(\theta)}
     {\inf_{\vartheta\in N_o}L_m(\vartheta)}
\frac{\Pi_O(N_o^c)}{\Pi_O(N_o)}
\le r_m.
\end{align}
Converting the odds bound to a probability bound proves~\eqref{eq:cocoon_bound}.
\end{proof}

The bound makes the intuitive competition explicit.
A concentrated bridge makes $\delta/(1-\delta)$ small, whereas the potential contribution of personalized target-domain evidence grows as $\exp(m\kappa)$.
When $m$ is small, the user's history may provide insufficient evidence to overcome the overlap concentration, leaving the inferred preference within $N_o$ with high probability.
The result does not imply permanent trapping: sufficiently informative target-domain evidence may overcome the transfer belief.

\subsection{Diversity Homogenization}
\label{subsec:homogenization}

To connect preference trapping to observable diversity, consider the standard transfer-dominance abstraction
\begin{equation}
    p_u=\lambda_u q_o+(1-\lambda_u)r_u,
    \qquad \lambda_u\in[0,1],
    \qquad u\in\U_n,
    \label{eq:transfer_mixture}
\end{equation}
where $p_u$ is the user's target recommendation distribution,
$q_o$ is the overlap-shaped component, $r_u$ is the target-only personalized component, and $\lambda_u$ is the effective transfer weight.
Theorem~\ref{thm:cocoon} motivates the corresponding regime in which the overlap-induced component can dominate.

\paragraph{Interpretation.}
Since deriving theoretical results for every backbone is impractical, we instead provide a conditional analytical model of transfer dominance. The overlap-induced belief, its high-mass region, and the transfer weights are analytical constructs that are not estimated for the evaluated backbones. The resulting bounds characterize a possible regime in which overlap-related influence limits diversity.

For distributions $p$ and $q$ on $\I_t$, let
$\TV(p,q)=\frac{1}{2}\sum_{i\in\I_t}|p(i)-q(i)|$.

\begin{corollary}[Inherited diversity and between-user collapse]
\label{cor:homogenization}
Suppose $\lambda_u\ge\lambda$ for all $u\in\U_n$.
Then
\begin{equation}
    \TV(p_u,q_o)\le 1-\lambda
    \label{eq:inheritance_tv}
\end{equation}
and, for any $u,v\in\U_n$,
\begin{equation}
    \TV(p_u,p_v)\le 2(1-\lambda).
    \label{eq:pairwise_tv}
\end{equation}
If all users share the same transfer weight $\lambda$, the latter sharpens to
$\TV(p_u,p_v)\le1-\lambda$.
Moreover, when $1-\lambda\le1-1/n$,
\begin{equation}
    |H(p_u)-H(q_o)|
    \le
    \Hb(1-\lambda)+(1-\lambda)\log(n-1).
    \label{eq:entropy_inheritance}
\end{equation}
\end{corollary}

\begin{proof}
Equation~\eqref{eq:transfer_mixture} gives
\[
\TV(p_u,q_o)=(1-\lambda_u)\TV(r_u,q_o)\le1-\lambda.
\]
The triangle inequality yields~\eqref{eq:pairwise_tv}; if the weights are equal, the shared $q_o$ term cancels.
Equation~\eqref{eq:entropy_inheritance} follows from the finite-alphabet continuity bound for entropy.
\end{proof}

Corollary~\ref{cor:homogenization} reveals two diversity failures at once.
Each non-overlapping user's distribution is constrained to lie near the bridge distribution, and different users lose distributional individuality as $\lambda$ approaches one.
This is the formal version of ``the bridge is too loud.''
The CDR cocoon is therefore not simply a low-entropy list: it is excessive similarity to the overlap bridge and to other non-overlapping users.

\subsection{Empirical Diagnosis}
\label{sec:empirical-diagnosis}
\begin{figure}[t]
    \centering
    \includegraphics[width=\columnwidth]{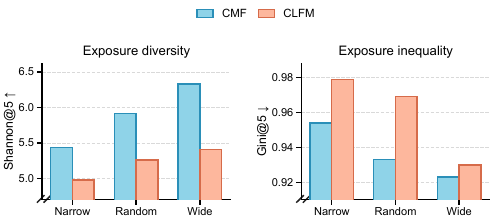}
    \caption{Effect of bridge width on diversity in Amazon.}
    \label{fig:bridge-width}
        \vspace{-4mm}
\end{figure}

\begin{figure}[t]
    \centering
    \includegraphics[width=\columnwidth]{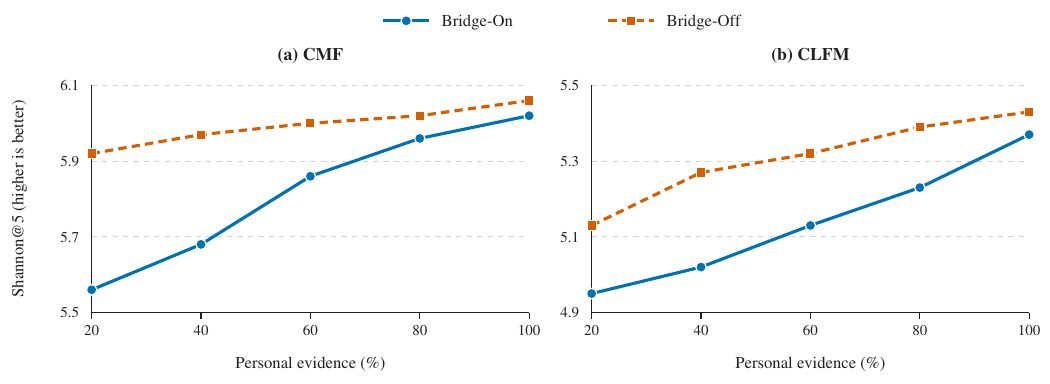}
    \caption{Evidence for the loud-bridge bottleneck in Amazon}
    \label{fig:target-evidence}
        \vspace{-4mm}
\end{figure}

We use two controlled experiments to illustrate the two bottlenecks.
\begin{itemize}
    \item \textbf{Bottleneck I.} We construct narrow, random, and wide groups of overlapping bridges based on their target-item coverage and then match users across strata on source-history length, target-history length, and mean popularity of their target items. As shown in Figure~\ref{fig:bridge-width}, widening the bridge increases diversity. A narrow bridge carries only a narrow range of tastes. These results demonstrate that bridge width is associated with recommendation diversity under this controlled construction.
    \item \textbf{Bottleneck II.} We conduct experiments by varying bridge availability (\textsc{On}/\textsc{Off}) and retaining $20\%$--$100\%$ of each user's target-domain history. \textsc{Bridge-Off} removes cross-domain identity links while preserving all observed interactions. Figure~\ref{fig:target-evidence} shows that enabling the cross-domain bridge consistently reduces recommendation diversity, with a generally larger reduction when personal target-domain evidence is limited. This interaction is consistent with the hypothesis that cross-domain signals exert greater influence on users with sparse personal evidence. 
\end{itemize}

%% file: 4.methodology.tex
\section{Methodology}
\label{sec:method}

\begin{figure*}
    \centering
    \includegraphics[width=0.85\linewidth]{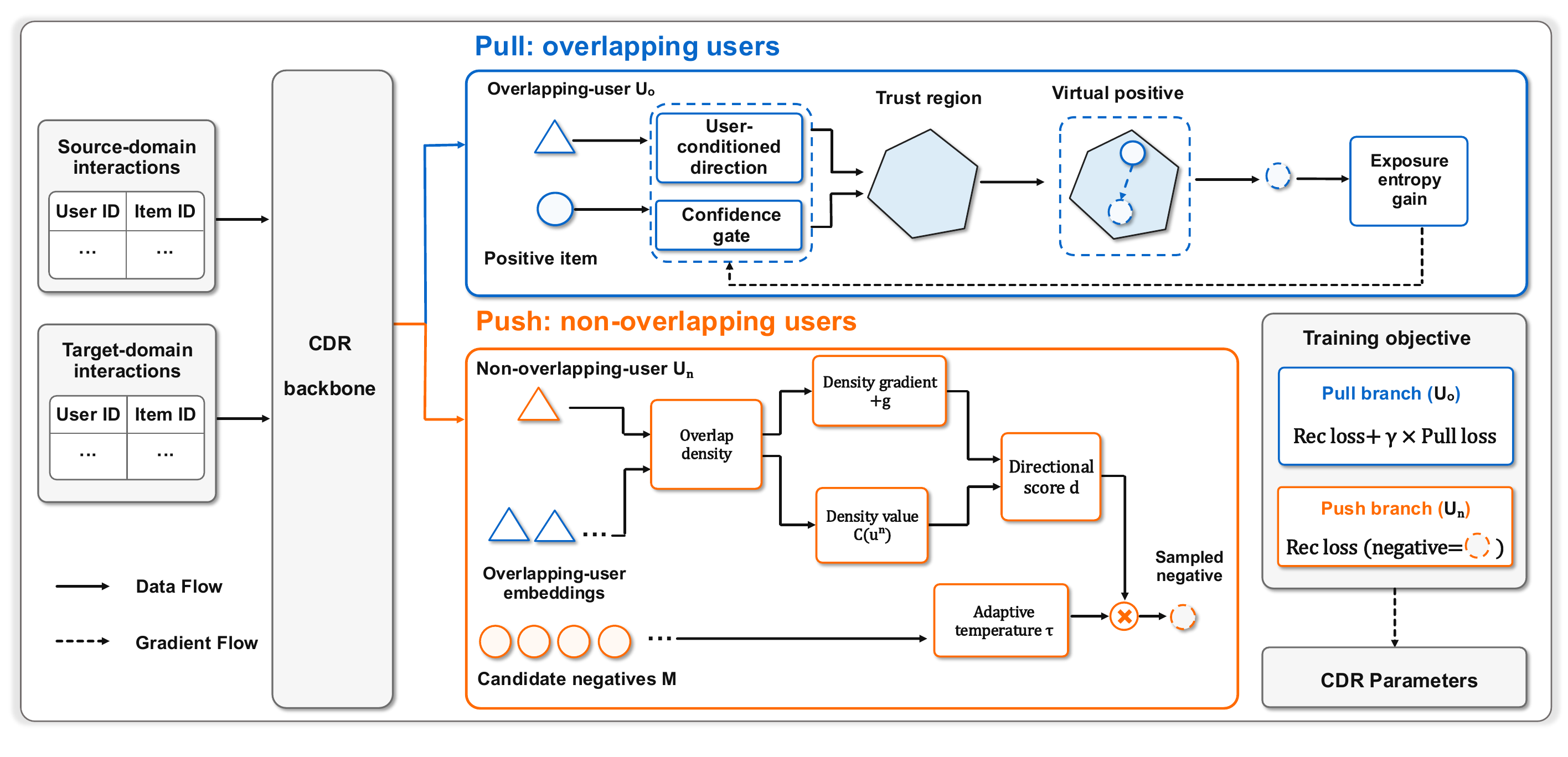}
    \vspace{-3mm}
    \caption{The illustration of our proposed DivCDR.}
    \label{fig:model_illustration}
        \vspace{-3mm}
\end{figure*}

Section~\ref{sec:theory} identifies two distinct bottlenecks to diversity in CDR. To address them, we propose \textbf{DivCDR}, a diversity-aware, model-agnostic Push--Pull framework for CDR. The overall workflow is depicted in Figure~\ref{fig:model_illustration}. 
Because our focus is the target domain, variables without an explicit $t$ (target) or $s$ (source) subscript refer to the target domain unless stated otherwise.

\subsection{From the Diagnosis to Two Design Requirements}
\label{subsec:design_requirements}
The analysis in Section~\ref{sec:theory} yields two design requirements.

\begin{itemize}
    \item \textbf{R1: broaden the bridge.} The entropy ceiling in Equation~\eqref{eq:bridge_entropy_bound} can be relaxed by moving overlap exposure beyond its concentrated support $S_o$. Thus, for $u^o\in\U_o$, an intervention should increase a differentiable exposure-entropy objective while bounding the perturbation of an observed positive-item embedding.
    \item \textbf{R2: escape overlap dominance.} Equations~\eqref{eq:cocoon_bound}--\eqref{eq:entropy_inheritance} indicate that a sparse $u^n\in\U_n$ needs a user-specific update away from the overlap-supported region.
\end{itemize}

To realize these two design requirements, we draw inspiration from the geometric dynamics of recommendation learning: positive samples attract a user representation (Pull), whereas negative samples repel it (Push)~\cite{RFG12,HYC17,ZZY22}. This observation leads to the following objectives:
\begin{align}
\text{(R1)}\quad
&\max_{\mathbf e_{i^+}'}
\ \Delta\mathcal H_{u^o}
\quad
\text{s.t.}\quad
\|\mathbf e_{i^+}'-\mathbf e_{i^+}\|_2
\leq
\alpha_{u_o,i^+}
\label{eq:requirement_pull}\\
\text{(R2)}\quad
&\min_{S_{\mathrm{push}}}
\ \E\!\left[
C\!\left(\mathbf e_{u^n}+\Delta_{u^n}^{-}(i^-)\right)
\mid u^n\right],
\quad i^-\sim S_{\mathrm{push}}(\cdot\mid u^n).
\label{eq:requirement_push}
\end{align}
R1 constructs a virtual positive $\mathbf e_{i^+}'$ that increases the overlap-side exposure entropy $\Delta\mathcal H_{u^o}$. The constraint keeps the virtual positive within a neighborhood of the observed positive $\mathbf e_{i^+}$, thereby bounding its displacement in embedding space. R2 selects a negative item $i^-$ through a sampling policy $S_{\mathrm{push}}$ such that the resulting sample reduces the expected cocoon strength $C$ of a non-overlapping user. We next instantiate these two requirements.
\subsection{Pull: Broadening the Overlap Bridge}
\label{subsec:pull}

The Pull module exploits the attractive effect of positive samples to broaden the preferences represented by overlapping users and thereby broaden the overlap bridge. Finding such positive samples is difficult: a suitable item must genuinely match the user's preference while also improving diversity. Because each user has only a limited number of observed interactions, an item satisfying both requirements may not exist among the available positive samples. Motivated by advances in data augmentation~\cite{YYX22,ZCL,ZCL23}, we therefore enhance observed positives by synthesizing new samples that jointly serve the accuracy and diversity objectives.

\paragraph{Step 1: Construct a user-conditioned virtual positive.}
For an observed positive item $i^+$ of an overlapping user $u^o$, we follow the augmentation idea in~\cite{YYX22} and perturb its embedding along a specified direction with a controlled scale:
\begin{align}
    \boldsymbol\epsilon_{u^o}
    &=   \frac{
\operatorname{MLP}_{\mathrm{aug}}(\mathbf e_{u^o})
}{
\left\|
\operatorname{MLP}_{\mathrm{aug}}(\mathbf e_{u^o})
\right\|_2
},
    \\
    \mathbf e_{i^+}'
    &=\mathbf e_{i^+} 
    +\alpha_{u^o,i^+}\boldsymbol \cdot \epsilon_{u^o}.
    \label{eq:virtual_positive}
\end{align}
Here, $\mathbf e_{i^+}$ denotes the original positive-item embedding. The first expression maps the overlapping-user representation to a personalized augmentation direction $\boldsymbol\epsilon_{u^o}$. The second moves the item only along this direction, using a magnitude $\alpha_{u^o,i^+}$.

However, this construction alone is insufficient. Without an appropriate supervisory signal, there is no guarantee that $\mathbf e_{i^+}'$ will satisfy the required accuracy and diversity properties. We therefore introduce an explicit objective and an adaptive constraint to supervise the generation process.

\paragraph{Step 2: Turn diversity into a differentiable exposure objective.}

Let $\mathcal B_I$ and $\mathcal B_U$ denote the candidate items and users in a mini-batch, respectively, and let $\mathcal B_U^o=\mathcal B_U\cap\U_o$. For a batch embedding set $\mathcal E$, we define the soft exposure distribution and its Shannon entropy for user $u$ as
\begin{align}
    \pi_u(j;\mathcal E)
    &=
    \frac{\exp(z_{uj})}
    {\sum_{\ell\in\mathcal B_I}\exp(z_{u\ell})},
    \label{eq:soft_exposure}\\
    \mathcal H_{\mathcal B}(u;\mathcal E)
    &=-\sum_{j\in\mathcal B_I}
    \pi_u(j;\mathcal E)\log\pi_u(j;\mathcal E),
    \label{eq:batch_entropy}
\end{align}
where $z_{uj}=f(\mathbf e_u,\mathbf e_j)$ is the backbone scoring function (e.g., an inner product). A larger $\mathcal H_{\mathcal B}(u;\mathcal E)$ corresponds to a more diffuse soft exposure distribution for user $u$. Then we use a similar approach to measure the aggregated exposure diversity of overlapping users within the mini-batch:
\begin{align}
\bar\pi_o(j;\mathcal E)
    &=\frac{1}{|\mathcal B_U^o|}
      \sum_{v\in\mathcal B_U^o}\pi_v(j;\mathcal E),
    \label{eq:group_exposure}\\
    \mathcal H_o(\mathcal E)
    &=-\sum_{j\in\mathcal B_I}
      \bar\pi_o(j;\mathcal E)\log\bar\pi_o(j;\mathcal E).
    \label{eq:group_exposure_entropy}
\end{align}
This $\mathcal H_o(\mathcal E)$ is important because an augmentation created for one user should not inadvertently reduce diversity for other users. Then, we measure the diversity improvement produced by the virtual positive relative to the original positive. Let $\mathcal E'$ be obtained from $\mathcal E$ by replacing $\mathbf e_{i^+}$ with $\mathbf e_{i^+}'$. We define
\begin{align}
    \Delta\mathcal H_{u^o}
    &=\mathcal H_{\mathcal B}(u^o;\mathcal E')
      -\mathcal H_{\mathcal B}(u^o;\mathcal E),
    \label{eq:focal_entropy_gain}\\
    \Delta\mathcal H_{\mathrm{group}}
    &=\mathcal H_o(\mathcal E')
      -\mathcal H_o(\mathcal E).
    \label{eq:group_entropy_gain}
\end{align}
Equation~\eqref{eq:focal_entropy_gain} quantifies how much the augmentation diversifies the exposure of the overlapping user. Equation~\eqref{eq:group_entropy_gain} measures the change in the aggregated exposure diversity of overlapping users.

We want the diversity gain after augmentation to be as large as possible. Accordingly, the augmentation parameters $\phi$ are trained by maximizing
\begin{equation}
    J(\phi)
    =\Delta\mathcal H_{u^o}
    +\Delta\mathcal H_{\mathrm{group}},
    \qquad
    \mathcal L_{\mathrm{aug}}=-J(\phi).
    \label{eq:augmentation_loss}
\end{equation}
Both terms are entropy differences; we assign them equal weight to avoid introducing an additional balancing hyperparameter.

\paragraph{Step 3: Learn a confidence-gated trust region.}
Step 2 determines how the direction $\boldsymbol\epsilon_{u^o}$ is learned. We must next determine the scale $\alpha_{u^o,i^+}$. Its purpose is to adaptively control the strength of augmentation. An excessively large perturbation may compromise the integrity of the original embedding and thereby impair accuracy. Conversely, an insufficient perturbation may prevent the generated embedding from effectively broadening the user's preference.

To choose a suitable augmentation strength, we draw inspiration from robustness analysis~\cite{GDH22,DQY20} and introduce the model's confidence in the observed positive interaction. Let
$\mathcal B_{I,u^o}^{-}=\mathcal B_I\setminus Y_{u^o}$ denote the unobserved candidate items for $u^o$ in the current mini-batch. We define the confidence margin of $i^+$ as
\begin{equation}
    m(u^o,i^+)
    =
    f(\mathbf e_{u^o},\mathbf e_{i^+})
    -\log\!\left(
      \frac{1}{|\mathcal B_{I,u^o}^{-}|}
      \sum_{j\in\mathcal B_{I,u^o}^{-}}
      \exp\!\left(
        f(\mathbf e_{u^o},\mathbf e_j)
      \right)
    \right).
    \label{eq:confidence_margin}
\end{equation}
The log-mean-exp term provides a smooth reference score for the unobserved candidates. A larger $m(u^o,i^+)$ indicates that the observed positive is more confidently separated from the candidate negatives, in which case a larger augmentation scale can be used. We therefore define:
\begin{equation}
    \alpha_{u^o,i^+}
    =\alpha_{\mathrm{base}}
    \operatorname{sigmoid}\!\left(
      m(u^o,i^+)
    \right)
    \operatorname{sigmoid}\!\left(
      \operatorname{MLP}_{\alpha}
      [\mathbf e_{u^o};\mathbf e_{i^+}]
    \right).
    \label{eq:adaptive_alpha}
\end{equation}
Here, hyperparameter $\alpha_{\mathrm{base}}>0$ serves as the maximum augmentation scale. The first sigmoid is a confidence gate, whereas the second remains a learned pair-specific gate. Because both gates lie in $(0,1)$, the augmentation scale satisfies
$0<\alpha_{u^o,i^+}<\alpha_{\mathrm{base}}$.

If $f(\mathbf e_{u^o},\cdot)$ is locally $L_f(u^o)$-Lipschitz, then
\[
\left|
f(\mathbf e_{u^o},\mathbf e'_{i^+})
-f(\mathbf e_{u^o},\mathbf e_{i^+})
\right|
\le
L_f(u^o)\alpha_{\mathrm{base}}
\|\boldsymbol\epsilon_{u^o}\|_2.
\]

\begin{proposition}[Local Pull improvement]
\label{prop:pull_ascent}
Suppose $J$ in Equation~\eqref{eq:augmentation_loss} is $L_J$-smooth in $\phi$.
For the ascent update $\phi^+=\phi+\eta\nabla J(\phi)$ with $0<\eta\le1/L_J$,
\begin{equation}
    J(\phi^+)
    \ge J(\phi)
    +\eta\left(1-\frac{L_J\eta}{2}\right)
     \|\nabla J(\phi)\|_2^2.
    \label{eq:pull_guarantee}
\end{equation}
The added term is non-negative; hence, this update cannot decrease $J$.
\end{proposition}

\begin{proof}
The smoothness ascent lemma yields
\begin{equation}
J(\phi^+)\ge J(\phi)
+\langle\nabla J(\phi),\phi^+-\phi\rangle
-\frac{L_J}{2}\|\phi^+-\phi\|_2^2.
\label{eq:pull_ascent_lemma}
\end{equation}
Equation~\eqref{eq:pull_ascent_lemma} lower-bounds the objective change by its first-order gain minus a curvature penalty.
Substituting $\phi^+-\phi=\eta\nabla J(\phi)$ gives Equation~\eqref{eq:pull_guarantee}.
\end{proof}

During this step, the augmentation parameters are updated using the
auxiliary diversity objective while the CDR backbone is held fixed. The proposition guarantees improvement in the implemented mini-batch entropy surrogate, rather than universal expansion along every embedding direction.

\subsection{Push: Weakening Overlap Dominance}
\label{subsec:push}
The Push module exploits the repulsive effect of negative samples to help non-overlapping users escape from regions dominated by overlapping users, thereby weakening the influence of an overly loud bridge.
To identify such negatives, Push first quantifies the user's cocoon strength, derives the corresponding escape direction, converts that direction into a directional alignment score for candidate negatives, and finally performs cocoon-adaptive sampling.

\paragraph{Step 1: Operationalize cocoon strength.}
We measure cocoon strength by the similarity between a non-overlapping user's soft exposure and the aggregated exposure of overlapping users:
\begin{equation}
    C(u^n)
    =1-\frac{
      \operatorname{JS}\!\left(
        \pi_{u^n}(\cdot;\mathcal E)
        \middle\|
        \operatorname{sg}[\bar\pi_o(\cdot;\mathcal E)]
      \right)}
      {\log 2},
    \qquad u^n\in\U_n,
    \label{eq:cocoon_strength}
\end{equation}
where $\operatorname{JS}$ is the Jensen--Shannon divergence computed with natural logarithms, and $\operatorname{sg}$ holds the group exposure fixed during a Push update. We write $C(\mathbf e_{u^n})$ when emphasizing its dependence on the user embedding. Thus, $C(u^n)\in[0,1]$, with a larger value indicating stronger exposure similarity to overlapping users. Moreover,
\begin{equation}
    D_{\mathrm{TV}}\!\left(
      \pi_{u^n}(\cdot;\mathcal E),
      \bar\pi_o(\cdot;\mathcal E)
    \right)
    \ge
    \frac{\operatorname{JS}(\pi_{u^n}\|\bar\pi_o)}{\log 2}
    =1-C(u^n).
    \label{eq:cocoon_tv_link}
\end{equation}
We reduce the cocoon strength along its gradient:
\begin{equation}
    \mathbf g_{u^n}
    :=\nabla_{\mathbf e_{u^n}}C(u^n).
    \label{eq:cocoon_gradient}
\end{equation}
The vector $\mathbf g_{u^n}$ points toward the steepest local increase in exposure similarity, so $-\mathbf g_{u^n}$ is the steepest local escape direction.

\paragraph{Step 2: Convert the escape direction into a directional negative score.}
Let $w_{u,i^-}$ be the derivative of the backbone loss with respect to the score of a negative item. For standard recommendation losses,
\begin{equation}
    w_{u,i^-}
    :=\frac{\partial\mathcal L_{\mathrm{rec}}}
    {\partial f(\mathbf e_u,\mathbf e_{i^-})}
    \ge0,
    \qquad
    \Delta_u^{-}(i^-)
    =-\eta_{\mathrm{rec}}w_{u,i^-}
      \nabla_{\mathbf e_u}
      f(\mathbf e_u,\mathbf e_{i^-}).
    \label{eq:negative_displacement}
\end{equation}
The first expression states that increasing a negative item's score increases the loss. The second isolates the contribution of that negative to the gradient-descent update of the user representation, explaining why a negative acts as a repulsive signal in the scoring geometry. In other words, because a negative sample supplies a repulsive signal, it should not itself be placed along the steepest escape direction. Instead, the ideal negative should lie on the $+\mathbf g_{u^n}$ side of the user. Repulsion from a negative in that region then moves the non-overlapping user toward $-\mathbf g_{u^n}$, thereby reducing $C(u^n)$ to first order. Based on this observation, we define the normalized cocoon gradient and the directional alignment score:
\begin{equation}
    \widehat{\mathbf g}_{u^n}
    =\frac{\mathbf g_{u^n}}
    {\|\mathbf g_{u^n}\|_2+\epsilon_0},
    \qquad
    d_{u^n}(i)
    =\left\langle
      \widehat{\mathbf g}_{u^n},
      \nabla_{\mathbf e_{u^n}}
      f(\mathbf e_{u^n},\mathbf e_i)
    \right\rangle,
    \label{eq:directional_score}
\end{equation}
where $\epsilon_0>0$ is a fixed numerical stabilizer. A candidate with $d_{u^n}(i)>0$ induces a negative update whose first-order cocoon-strength change is non-positive, because Equations~\eqref{eq:negative_displacement} and~\eqref{eq:directional_score} give
\begin{equation}
    \left\langle
      \mathbf g_{u^n},
      \Delta_{u^n}^{-}(i)
    \right\rangle
    =-\eta_{\mathrm{rec}}w_{u^n,i}
      \bigl(\|\mathbf g_{u^n}\|_2+\epsilon_0\bigr)
      d_{u^n}(i)
    \le 0.
    \label{eq:directional_descent}
\end{equation}
For an inner-product scorer, $d_{u^n}(i)$ reduces to
$\langle\widehat{\mathbf g}_{u^n},\mathbf e_i\rangle$. Thus, the score is inexpensive to compute and directly measures whether repulsion from a candidate item agrees with the escape direction.

\paragraph{Step 3: Sample a cocoon-adaptive negative.}
Given a random candidate pool $M\subseteq\I_t\setminus Y_{u^n}$, let
$M_{u^n}^{+}=\{i\in M:d_{u^n}(i)>0\}$ denote the candidates whose repulsive updates agree with the local escape direction. When $M_{u^n}^{+}$ is non-empty, we sample a real negative according to
\begin{equation}
    p(i\mid u^n)
    =\frac{
    \exp\!\left(
      d_{u^n}(i)/\tau(u^n)
    \right)}{
    \sum_{j\in M_{u^n}^{+}}
    \exp\!\left(
      d_{u^n}(j)/\tau(u^n)
    \right)},
    \qquad i\in M_{u^n}^{+}.
    \label{eq:push_sampling}
\end{equation}
If $M_{u^n}^{+}$ is empty, Push falls back to the backbone's original negative sampler for that update. The descent characterization below applies to Push updates sampled from a non-empty $M_{u^n}^{+}$.
The candidate pool $M$ provides two advantages. First, computing negative-sampling probabilities over the complete set of unobserved items would be prohibitively expensive. Second, drawing a smaller candidate subset at random limits the opportunity to select an unobserved but genuinely preferred item as a negative, under the standard assumption that such items account for only a small fraction of the candidate pool. The temperature $\tau(u^n)$ adapts the sampling distribution to the user's cocoon strength. A larger $C(u^n)$ indicates that the user is more tightly confined within the cocoon and therefore requires a lower temperature, which produces a sharper sampling distribution. This sharper distribution favors negatives with larger directional alignment scores, providing a stronger signal for diversity enhancement. Accordingly, we define
\begin{equation}
    \tau(u^n)=\exp(-C(u^n)).
    \label{eq:adaptive_temperature}
\end{equation}

\begin{proposition}[Cocoon-adaptive directional hardness]
\label{prop:temperature_hardness}
Let $s_i=d_{u^n}(i)$ and
$p_\beta(i)=\exp(\beta s_i)/\sum_{j\in M_{u^n}^{+}}\exp(\beta s_j)$
for a non-empty $M_{u^n}^{+}$.
Then
\begin{equation}
    \frac{\partial}{\partial\beta}
    \E_{i\sim p_\beta}[s_i]
    =\operatorname{Var}_{i\sim p_\beta}(s_i)
    \ge0.
    \label{eq:temperature_variance}
\end{equation}
Thus, a larger $C(u^n)$ produces a larger inverse temperature $\beta(u^n)=1/\tau(u^n)$ and cannot decrease the expected directional alignment score of the sampled negative.
\end{proposition}

\begin{proof}
Differentiating the log-partition function gives
\begin{equation}
    \frac{\partial}{\partial\beta}
    \log\sum_{j\in M_{u^n}^{+}}e^{\beta s_j}
    =\E_{i\sim p_\beta}[s_i],
    \qquad
    \frac{\partial^2}{\partial\beta^2}
    \log\sum_{j\in M_{u^n}^{+}}e^{\beta s_j}
    =\operatorname{Var}_{i\sim p_\beta}(s_i).
    \label{eq:log_partition}
\end{equation}
The first equality identifies expected hardness as the first derivative of the log-partition function, whereas the second identifies its rate of change as a non-negative variance, proving Equation~\eqref{eq:temperature_variance}.
\end{proof}

Proposition~\ref{prop:temperature_hardness} formalizes the temperature intuition: among candidates whose repulsive updates agree with the escape direction, a strongly cocooned user is assigned a more directionally aligned negative, whereas a weakly cocooned user retains a softer and more conservative sampler.

\paragraph{Step 4: Characterize when sampled Push decreases cocoon strength.}
The ideal gradient direction is continuous, whereas the implementation can act only through a sampled catalog item. The next result makes this approximation explicit rather than treating the directional score as an exact optimizer update.

\begin{theorem}[Expected Push descent]
\label{thm:push_descent}
Assume $C$ is $L_C$-smooth around $\mathbf e_{u^n}$, and let $\Delta_{u^n}^{-}$ be the negative-induced displacement from Equation~\eqref{eq:negative_displacement}, with $i^-\sim p(\cdot\mid u^n)$.
Suppose $M_{u^n}^{+}$ is non-empty and define
\begin{align}
    A_{u^n}
    &:=\eta_{\mathrm{rec}}
      \bigl(\|\mathbf g_{u^n}\|_2+\epsilon_0\bigr)
      \E\!\left[
        w_{u^n,i^-}d_{u^n}(i^-)
        \mid u^n
      \right]\ge0,
          \label{eq:stochastic_push_conditions}
    \\
    B_{u^n}
    &:=\E\!\left[
      \|\Delta_{u^n}^{-}\|_2^2
      \mid u^n
    \right]\ge0.
    \label{eq:stochastic_push_conditions2}
\end{align}
Then
\begin{equation}
    \E\!\left[
      C(\mathbf e_{u^n}+\Delta_{u^n}^{-})
      \mid u^n
    \right]
    \le C(\mathbf e_{u^n})
    -A_{u^n}+\frac{L_C}{2}B_{u^n}.
    \label{eq:stochastic_push_bound}
\end{equation}
In particular, the expected cocoon strength does not increase whenever
$A_{u^n}\ge L_CB_{u^n}/2$.
\end{theorem}

Combining Equations~\eqref{eq:cocoon_tv_link} and~\eqref{eq:stochastic_push_bound} gives
\begin{equation}
    \E\!\left[
      D_{\mathrm{TV}}\!\left(
        \pi_{u^n}^{+},\bar\pi_o
      \right)
      \middle|u^n
    \right]
    \ge 1-C(u^n)
      +A_{u^n}-\frac{L_C}{2}B_{u^n},
    \label{eq:push_tv_bound}
\end{equation}
where $\pi_{u^n}^{+}$ is the soft exposure after the sampled Push update.

Equation~\eqref{eq:directional_descent} guarantees that every negative sampled from $M_{u^n}^{+}$ has non-positive first-order cocoon-strength change. Accordingly, $A_{u^n}$ in Equation~\eqref{eq:stochastic_push_conditions} is the expected first-order decrease induced by the sampler, rather than a separately imposed alignment assumption. The quantity $B_{u^n}$ measures the second moment of the update. Equation~\eqref{eq:stochastic_push_bound} shows that descent occurs when the sampler's directional progress dominates the curvature penalty.

\begin{proof}
$L_C$-smoothness gives, for every sampled displacement,
\begin{equation}
    C(\mathbf e+\Delta)
    \le C(\mathbf e)
    +\langle\nabla C(\mathbf e),\Delta\rangle
    +\frac{L_C}{2}\|\Delta\|_2^2.
    \label{eq:push_smoothness}
\end{equation}
Equation~\eqref{eq:push_smoothness} decomposes the cocoon-strength change into a directional first-order term and a quadratic curvature cost. By Equation~\eqref{eq:directional_descent} and the definitions,
$\E[\langle\mathbf g_{u^n},\Delta_{u^n}^{-}\rangle\mid u^n]=-A_{u^n}$ and
$\E[\|\Delta_{u^n}^{-}\|_2^2\mid u^n]=B_{u^n}$. Substitution into Equation~\eqref{eq:push_smoothness} proves Equation~\eqref{eq:stochastic_push_bound}.
\end{proof}
This theorem establishes the expected local effect induced by the
sampled negative, providing the optimization rationale for the Push
sampler within the complete recommendation objective.

For the ideal displacement $\Delta_{u^n}^{-}=-\eta\mathbf g_{u^n}$, the theorem reduces to
\begin{equation}
    C(\mathbf e_{u^n}-\eta\mathbf g_{u^n})
    \le C(\mathbf e_{u^n})
    -\eta\left(1-\frac{L_C\eta}{2}\right)
     \|\mathbf g_{u^n}\|_2^2,
    \qquad 0<\eta\le\frac{2}{L_C}.
    \label{eq:ideal_push_guarantee}
\end{equation}
The coefficient is non-negative within the stated step-size range, so exact gradient descent monotonically reduces the cocoon strength. In the implemented sampler, restricting the candidates to $M_{u^n}^{+}$ guarantees first-order alignment using available catalog items, while the Boltzmann distribution assigns larger probability to candidates with stronger alignment. Theorem~\ref{thm:push_descent} states the remaining step-size condition under which this discrete approximation inherits the expected descent property.

\subsection{Overall Objective}
\label{subsec:overall_training}

Let $S_{\mathrm{push}}(M,u^n)$ denote a negative drawn from Equation~\eqref{eq:push_sampling}. For target-domain instances, DivCDR uses the piecewise objective
\begin{equation}
\mathcal L_{\mathrm{DivCDR}}(u)=
\begin{cases}
\mathcal L_{\mathrm{rec}}
(\mathbf e_{u^o},\mathbf e_{i^+}',\mathbf e_{i^-})
+\gamma\mathcal L_{\mathrm{aug}},
&u\in\U_o,\\[1.5mm]
\mathcal L_{\mathrm{rec}}
(\mathbf e_{u^n},\mathbf e_{i^+},
 \mathbf e_{S_{\mathrm{push}}(M,u^n)}),
&u\in\U_n.
\end{cases}
\label{eq:overall_objective}
\end{equation}
The first branch replaces an overlapping user's observed positive embedding with its Pull augmentation and weights the entropy-gain loss by $\gamma$. The second keeps the positive of a non-overlapping user unchanged and replaces only the ordinary negative sampler with Push. DivCDR therefore requires access only to the backbone embeddings, scoring function, and recommendation loss; it does not modify the backbone architecture.

%% file: 5.experiment.tex
\section{Experiments}
\label{experiment}
In this section, we conduct experiments to answer the following questions: \textbf{RQ1:} \textit{What are the effects of DivCDR on the diversity and accuracy?} \textbf{RQ2:} \textit{How does DivCDR perform in comparison to state-of-the-art (SOTA) diversity-aware CDR methods?} \textbf{RQ3:} \textit{How does DivCDR fare under a range of widely adopted diversity evaluation metrics?} 
\textbf{RQ4:} \textit{How do different components and parameter settings of DivCDR influence its overall performance? }

\subsection{Experimental Setup}

\begin{table}
\footnotesize
\centering
\caption{Statistics of used datasets.}
\label{tab:statistics}
\begin{tabular}{llrrrr}
\toprule
\textbf{Dataset} & \textbf{Domain} & \textbf{\#Users} & \textbf{\#Items} & \textbf{\#Inter.} & \textbf{Overlap} \\
\midrule
\multirow{2}{*}{Amazon} & Book & 65,725 & 70,071 & 2,021,443 & 2.55\% \\
 & Movie & 8,663 & 7,790 & 229,257 & 19.32\% \\
\midrule
\multirow{2}{*}{Epinions} & Elec & 10,124 & 13,018 & 34,859 & 12.51\% \\
 & Game & 4,247 & 4,094 & 16,471 & 29.83\% \\
 \midrule
\multirow{2}{*}{TenRec} & Video & 8,682 & 65,216 & 539,193 & 21.78\% \\
 & Article & 21,639 & 1,859 & 93,832 & 8.74\% \\
\bottomrule
\end{tabular}
    \vspace{-3mm}
\end{table}

\textbf{Datasets.} We evaluate DivCDR on three widely used benchmark datasets in CDR research: \textbf{Amazon}, \textbf{TenRec}~\cite{yuan2022tenrec}, and \textbf{Epinions}. The detailed dataset statistics are summarized in Table~\ref{tab:statistics}.

\noindent \textbf{Evaluation Protocols.}  
To evaluate accuracy for top-K recommendations and full ranking, we include Hit@K and NDCG@K. For measuring diversity, we leverage Shannon Entropy (Shannon@K) and the Gini Index (Gini@K). While numerous other diversity metrics have been proposed for recommendation~\cite{kaminskas2016diversity,kunaver2017diversity}, we defer their discussion to Sec.~\ref{sec:metric} due to space constraints.

\noindent \textbf{Baselines.}  
Since DivCDR is model-agnostic, we evaluate it on representative CDR models~\cite{zhao2026double, zhao2026sc}: \textbf{CMF}~\cite{SG08}, \textbf{CLFM}~\cite{GLC13}, \textbf{BiTGCF}~\cite{LLL20}, \textbf{EMCDR}~\cite{MSJ17}, and \textbf{DTCDR}~\cite{zhu2019dtcdr}. Additionally, we compare DivCDR against diversity-related CDR algorithms, including \textbf{CCDR}~\cite{XLW22}, \textbf{AD-CDMF}~\cite{sun2022prick}, and \textbf{VUG}~\cite{chen2025leave}. 
Although \textbf{VUG} does not directly mention diversity, its idea of transforming all users into overlapping users may help improve diversity.

\noindent \textbf{Experimental Details.}  
To improve reproducibility, we implement DivCDR based on the RecBole CDR framework~\cite{ZMH21, ZHP22}, following its data pre-processing procedures, dataset splits, and evaluation protocol. All hyperparameters are thoroughly tuned using the respective validation sets. For our hyperparameter settings, the size of $M$ is selected from \{2, 4, 8, 16\}, while both the $\alpha_{base}$ and $\gamma$ are chosen within the range (0, 0.2].

\subsection{Overall Performance}
\begin{table*}[t]
\caption{Results of different CDR models with (w) or without (w/o) DivCDR on three datasets. The best results in each pair (w/o vs. w) are \textbf{boldfaced}.}
\centering
{
\begin{tabular}{@{}cl*{10}{l}@{}}
\toprule
\multirow{2}{*}{\textbf{Datasets}} & \multirow{2}{*}{\textbf{Metric}} & \multicolumn{2}{c}{\textbf{CMF}} & \multicolumn{2}{c}{\textbf{CLFM}} & \multicolumn{2}{c}{\textbf{BiTGCF}} & \multicolumn{2}{c}{\textbf{EMCDR}} & \multicolumn{2}{c}{\textbf{DTCDR}} \\ \cmidrule(l){3-12}
 &  & w/o & w & w/o & w & w/o & w & w/o & w & w/o & w \\ \midrule
\multirow{8}{*}{Amazon}
 & NDCG@5 & 0.0269 & \textbf{0.0301} & 0.0151 & \textbf{0.0152} & 0.0247 & \textbf{0.0279} & 0.0237 & \textbf{0.0251} & 0.0157 & \textbf{0.0184} \\
 & NDCG@10 & 0.0344 & \textbf{0.0379} & 0.0195 & \textbf{0.0198} & 0.0324 & \textbf{0.0366} & 0.0298 & \textbf{0.0316} & 0.0206 & \textbf{0.0239} \\
 & Hit@5 & 0.0630 & \textbf{0.0695} & \textbf{0.0383} & 0.0372 & 0.0584 & \textbf{0.0661} & 0.0510 & \textbf{0.0511} & 0.0387 & \textbf{0.0448} \\
 & Hit@10 & 0.1016 & \textbf{0.1031} & \textbf{0.0638} & 0.0627 & 0.0979 & \textbf{0.1025} & 0.0791 & \textbf{0.0793} & 0.0661 & \textbf{0.0682} \\
\cmidrule(l){2-12}
 & Shannon@5 & 6.0226 & \textbf{6.5315} & 5.3269 & \textbf{6.5831} & 6.5527 & \textbf{6.7260} & 7.0416 & \textbf{7.3811} & 5.1716 & \textbf{6.0239} \\
 & Shannon@10 & 6.3734 & \textbf{6.7731} & 5.8882 & \textbf{6.9411} & 6.7956 & \textbf{6.9452} & 7.3186 & \textbf{7.5824} & 5.7754 & \textbf{6.5169} \\
 & Gini@5 & 0.9498 & \textbf{0.9189} & 0.9747 & \textbf{0.9164} & 0.9184 & \textbf{0.9072} & 0.8504 & \textbf{0.7899} & 0.9693 & \textbf{0.9301} \\
 & Gini@10 & 0.9322 & \textbf{0.8997} & 0.9590 & \textbf{0.8852} & 0.8990 & \textbf{0.8841} & 0.8139 & \textbf{0.7524} & 0.9505 & \textbf{0.8980} \\
\midrule
\multirow{8}{*}{Epinions}
 & NDCG@5 & 0.0455 & \textbf{0.0511} & \textbf{0.0348} & 0.0344 & 0.0486 & \textbf{0.0499} & 0.0253 & \textbf{0.0257} & 0.0410 & \textbf{0.0420} \\
 & NDCG@10 & 0.0549 & \textbf{0.0591} & \textbf{0.0458} & 0.0447 & 0.0596 & \textbf{0.0601} & 0.0317 & \textbf{0.0329} & \textbf{0.0570} & 0.0561 \\
 & Hit@5 & 0.0664 & \textbf{0.0748} & 0.0557 & \textbf{0.0559} & 0.0703 & \textbf{0.0738} & 0.0351 & \textbf{0.0367} & 0.0653 & \textbf{0.0670} \\
 & Hit@10 & 0.0966 & \textbf{0.1033} & \textbf{0.0916} & 0.0877 & 0.1054 & \textbf{0.1086} & 0.0550 & \textbf{0.0631} & \textbf{0.1153} & 0.1127 \\
\cmidrule(l){2-12}
 & Shannon@5 & 4.5850 & \textbf{5.0450} & 3.7926 & \textbf{3.9301} & 4.2127 & \textbf{4.9355} & 5.9042 & \textbf{6.0349} & 1.7441 & \textbf{1.8375} \\
 & Shannon@10 & 4.9332 & \textbf{5.3666} & 4.2977 & \textbf{4.5176} & 4.5758 & \textbf{5.2780} & 6.1608 & \textbf{6.2668} & 2.4253 & \textbf{2.5101} \\
 & Gini@5 & 0.9749 & \textbf{0.9638} & 0.9900 & \textbf{0.9879} & 0.9816 & \textbf{0.9644} & 0.9322 & \textbf{0.9167} & 0.9987 & \textbf{0.9983} \\
 & Gini@10 & 0.9645 & \textbf{0.9484} & 0.9834 & \textbf{0.9805} & 0.9736 & \textbf{0.9489} & 0.9121 & \textbf{0.8947} & 0.9975 & \textbf{0.9969} \\
\midrule
\multirow{8}{*}{TenRec}
 & NDCG@5 & 0.0919 & \textbf{0.0998} & 0.0755 & \textbf{0.1031} & \textbf{0.1074} & 0.1059 & 0.0820 & \textbf{0.0971} & 0.0827 & \textbf{0.0948} \\
 & NDCG@10 & 0.1241 & \textbf{0.1332} & 0.1023 & \textbf{0.1359} & \textbf{0.1398} & 0.1367 & 0.1110 & \textbf{0.1277} & 0.1076 & \textbf{0.1261} \\
 & Hit@5 & 0.1480 & \textbf{0.1587} & 0.1229 & \textbf{0.1618} & \textbf{0.1682} & 0.1656 & 0.1319 & \textbf{0.1531} & 0.1347 & \textbf{0.1523} \\
 & Hit@10 & 0.2483 & \textbf{0.2561} & 0.2064 & \textbf{0.2656} & \textbf{0.2697} & 0.2591 & 0.2229 & \textbf{0.2498} & 0.2134 & \textbf{0.2481} \\
\cmidrule(l){2-12}
 & Shannon@5 & 3.1894 & \textbf{3.2269} & 2.5752 & \textbf{3.8108} & 3.4346 & \textbf{3.6733} & 4.2731 & \textbf{4.3134} & 1.8297 & \textbf{3.2445} \\
 & Shannon@10 & 3.6186 & \textbf{3.6335} & 3.0694 & \textbf{4.2500} & 3.9226 & \textbf{4.0115} & 4.5357 & \textbf{4.5637} & 2.4606 & \textbf{3.8467} \\
 & Gini@5 & 0.9897 & \textbf{0.9890} & 0.9929 & \textbf{0.9781} & 0.9857 & \textbf{0.9828} & 0.9680 & \textbf{0.9510} & 0.9970 & \textbf{0.9845} \\
 & Gini@10 & 0.9844 & \textbf{0.9836} & 0.9890 & \textbf{0.9687} & 0.9786 & \textbf{0.9771} & 0.9590 & \textbf{0.9420} & 0.9942 & \textbf{0.9746} \\
\bottomrule
\end{tabular}
}
\label{tab:main_results}
    \vspace{-3mm}
\end{table*}

Table~\ref{tab:main_results} reports results averaged over ten independent runs. Under the two-sided t-test used in our evaluation, the reported improvements have p < 0.05.
\begin{itemize}
    \item \textbf{Substantial Gains in Diversity:} Integrating DivCDR with various backbone models consistently yields improvements in diversity-related metrics. Most notably, integrating DivCDR with DTCDR improves Shannon@5 by 77.32\% on TenRec. These results underscore the effectiveness of DivCDR in improving diversity.
    \item \textbf{Preserved or Enhanced Accuracy:} DivCDR preserves or improves recommendation accuracy in most settings. The largest relative gain is 36.56\% for CLFM on TenRec; among settings with lower accuracy, the mean relative decrease is 2.30\%. Considering the longstanding trade-off between diversity and accuracy in recommendation systems, achieving such accuracy while significantly increasing diversity is truly remarkable.
    \item \textbf{Limitations of Traditional Methods:} In contrast, traditional CDR approaches demonstrate limited improvements in diversity despite rapid advances in accuracy. This shortcoming can detrimentally impact user satisfaction and exacerbate filter bubble effects, emphasizing the need for diversity-aware solutions in the ongoing evolution of cross-domain recommendation.
\end{itemize}

\subsection{Performance Comparison with Diversity-related CDR Methods}
\label{sec:metric}
\begin{figure}
    \centering
    \includegraphics[width=0.85\linewidth]{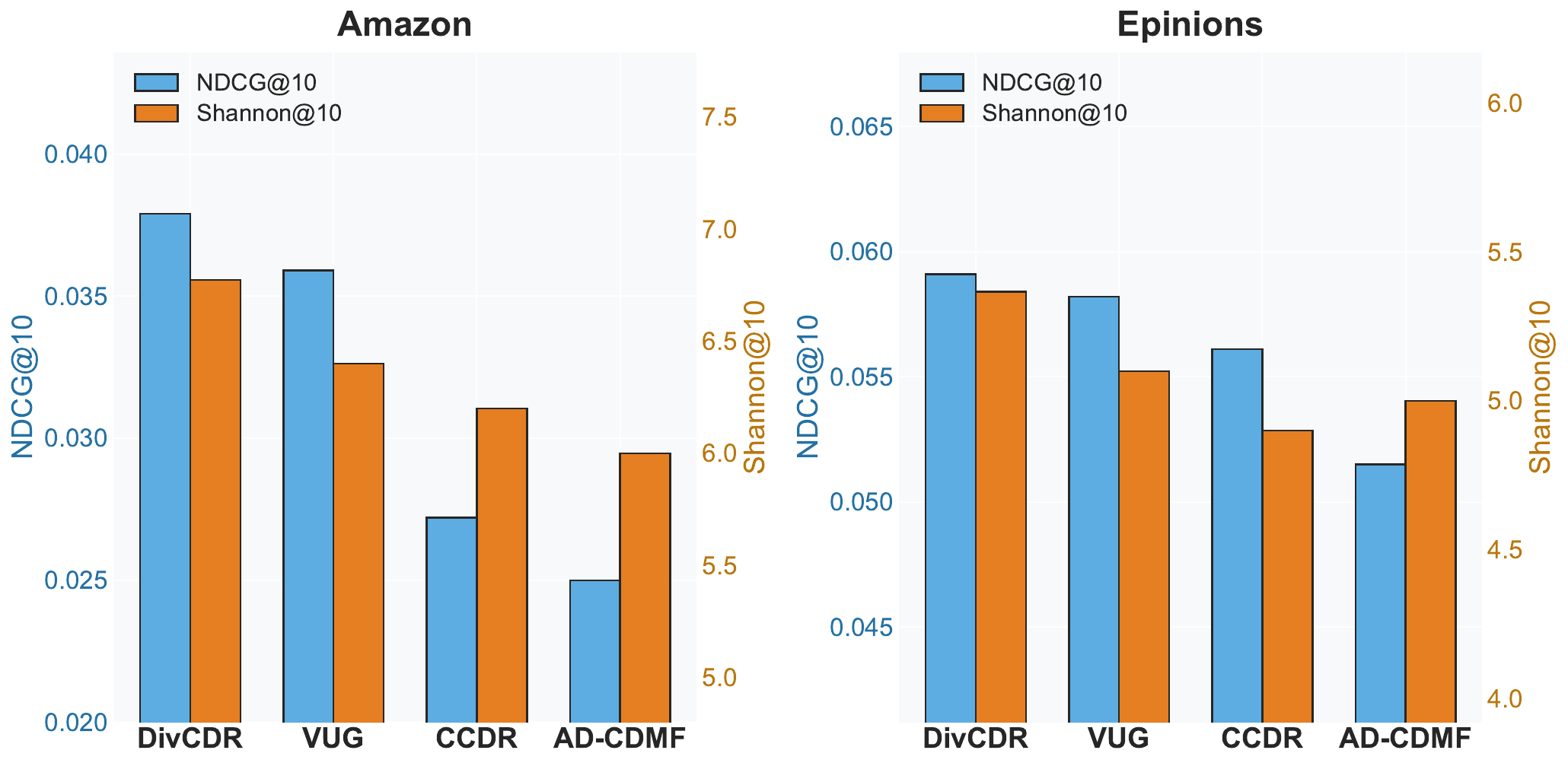}
    \caption{Comparison with diversity-related baselines.}
    \label{fig:baseline}
        \vspace{-4mm}
\end{figure}
To more comprehensively demonstrate the capabilities of DivCDR, we compared it against diversity-related CDR methods available in the literature as far as we know.
As shown in Figure~\ref{fig:baseline}, DivCDR consistently outperforms existing diversity-related CDR methods in both accuracy and diversity metrics. While CCDR and AD-CDMF both emphasize diversity enhancement, they primarily focus on directly increasing diversity without addressing the specific diversity constraints inherent to CDR. In contrast, our approach first conducts a thorough analysis of the diversity in CDR and subsequently proposes a targeted solution, thereby tackling the underlying issue and achieving superior performance.

\subsection{Additional Popular Diversity Metrics}
\label{sec:metric}
\begin{figure}
    \centering
    \includegraphics[width=0.85\linewidth]{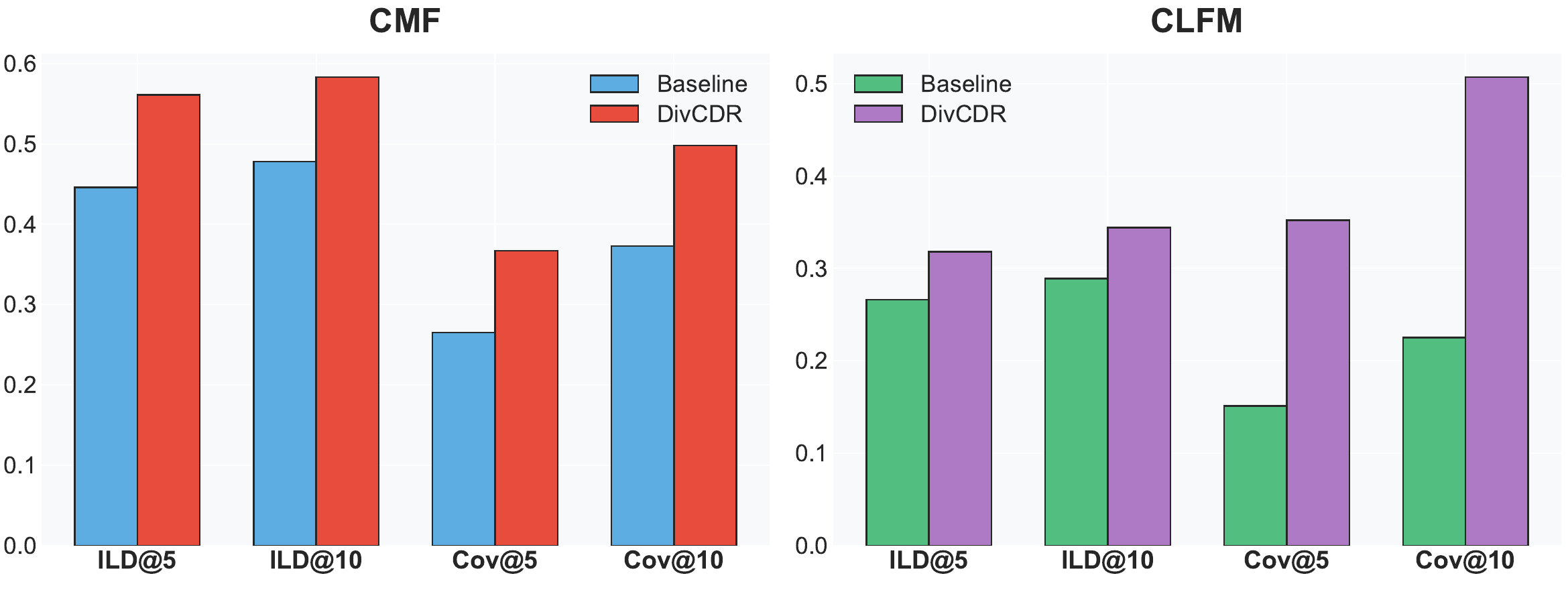}
    \caption{Diversity performance in terms of \textit{intra-list diversity} and \textit{item coverage} on the Amazon dataset.}
    \label{fig:metric}
        \vspace{-4mm}
\end{figure}

As previously noted, diversity can be assessed using a wide range of metrics, each with its own theoretical underpinnings~\cite{kaminskas2016diversity,kunaver2017diversity}. 
Consequently, it is challenging to provide an exhaustive discussion of all diversity metrics within a single paper. 
In addition to our main experimental results, we further evaluate DivCDR on two additional popular metrics, \textit{ILD} and \textit{Item Coverage}~\cite{wu2024result}, with the results summarized in Figure~\ref{fig:metric}. DivCDR continues to exhibit strong performance under these metrics.

\subsection{Ablation Study}
\begin{table}[t]
\footnotesize
\centering
\caption{Ablation study integrated with BiTGCF on the Epinions dataset.}
\label{tab:ablation}
{
\begin{tabular}{lcccc}
\toprule
\multirow{2}{*}{\textbf{Method}} & \multicolumn{4}{c}{\textbf{Accuracy}}\\
\cmidrule(lr){2-5}
& Hit@5 & Hit@10 & NDCG@5 & NDCG@10 \\ 
\midrule
\emph{w/o aug} & 0.0729 & 0.1054 & 0.0478 & 0.0577 \\ 
\emph{w/o push} & \textbf{0.0751} & \textbf{0.1089} & 0.0481 & 0.0589 \\ 
\emph{w/o pull} & 0.0649 & 0.1019 & 0.0430 & 0.0549 \\ 
\emph{w/o neg} & 0.0657 & 0.1042 & 0.0459 & 0.0577 \\
\textbf{DivCDR} & 0.0738 & 0.1086 & \textbf{0.0499} & \textbf{0.0601} \\ 
\midrule
\multirow{2}{*}{\textbf{Method}} & \multicolumn{4}{c}{\textbf{Diversity}}\\
\cmidrule(lr){2-5}
& Shannon@5 & Shannon@10 & Gini@5 & Gini@10 \\
\midrule
\emph{w/o aug} & 4.5089 & 4.8874 & 0.9771 & 0.9661 \\
\emph{w/o push} & 4.6077 & 4.9873 & 0.9714 & 0.9582 \\
\emph{w/o pull} & 4.6560 & 4.9831 & 0.9717 & 0.9610 \\
\emph{w/o neg} & 4.5416  & 4.9930 & 0.9769 & 0.9673 \\
DivCDR & \textbf{4.9355} & \textbf{5.2780} & \textbf{0.9644} & \textbf{0.9489} \\
\bottomrule
\end{tabular}
}
    \vspace{-3mm}
\end{table}

To evaluate the effect of different components, we conduct experiments on four variants: (i) DivCDR without the pulling strategy (w/o \textit{pull}), (ii) DivCDR without the pushing strategy (w/o \textit{push}), (iii) DivCDR without diversity loss (w/o \textit{aug}), and (iv) DivCDR with random negative sampling instead of the original sampling strategy (w/o \textit{neg}). Results are summarized in Table~\ref{tab:ablation}. The ablation results show that removing each component reduces the diversity performance. In terms of accuracy, we observe notable fluctuations when DivCDR operates without the pushing strategy. This outcome is expected, as the pushing mechanism primarily aims to help users break out of cocoons rather than to optimize accuracy.

\subsection{Hyperparameter Analysis}
\begin{figure}
    \centering
    \includegraphics[width=\linewidth]{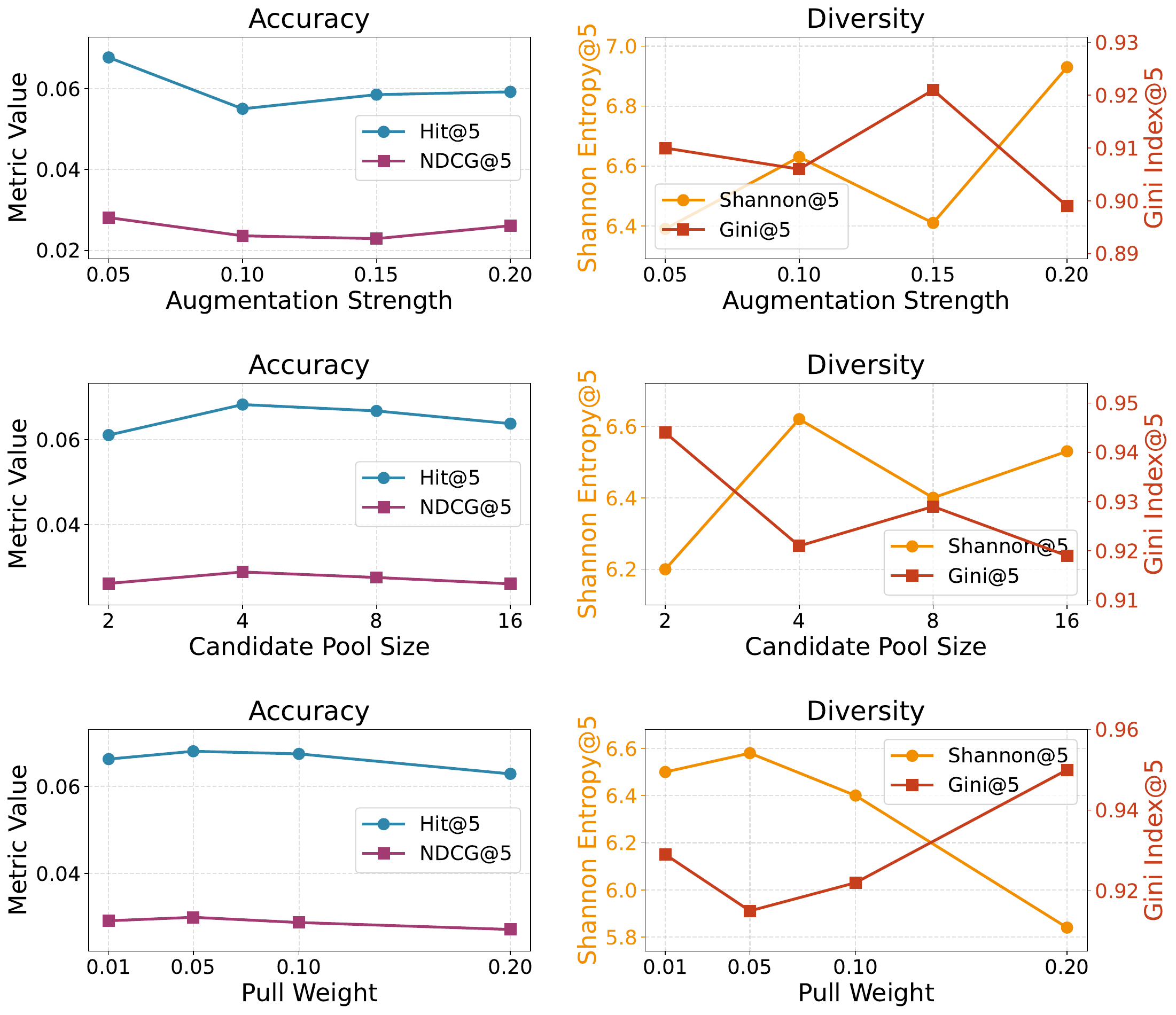}
    \caption{Hyperparameter sensitivity analysis.
    }
    \label{fig:hyper}
    \vspace{-3mm}
\end{figure}

We analyze the influence of three key hyperparameters in DivCDR: $\alpha_{\text{base}}$, $M$, and $\gamma$. Experimental results are depicted in Figure~\ref{fig:hyper}. Two key observations emerge: (i) Within reasonable ranges, all hyperparameters allow DivCDR to maintain stable accuracy and diversity performance. (ii)  Accuracy initially increases with larger $M$ and subsequently decreases, which is reasonable: with a larger candidate set, it becomes easier to select items that address user cocoons, but an excessively large set may lead to the selection of items that the user actually prefers, thereby harming recommendation accuracy.

%% file: 2.related_work.tex
\section{Related Work}

\subsection{Cross-Domain Recommendation}

The goal of cross-domain recommendation (CDR) is to enhance recommendations in a target domain by transferring information from a source domain. CMF~\cite{SG08} was among the earliest approaches to jointly factorize multiple rating matrices for knowledge transfer. 
PPCDR~\cite{tian2024privacy} adopted a federated graph learning framework to capture user preferences while preserving privacy. GDCCDR~\cite{liu2024graph} achieved personalized transfer by integrating meta-networks with contrastive learning. PCDR~\cite{guo2024prompt} utilized users' interactions on local clients and leveraged natural language to bridge domains. MACDR~\cite{WYW24} explored unsupervised methods for transferring non-overlapping knowledge. CD-CDR~\cite{CDCDR} employed a shared conditional diffusion model to learn unified item distributions, thereby facilitating cross-domain knowledge transfer. MF-GSLAE~\cite{WYZ25} employed a bootstrapping-based factor selection module to identify preferences unique to each domain. Lasso~\cite{chen2025lasso} presented a cross-domain training paradigm that fine-tuned an LLM-based simulator. 

\subsection{Diversity in Cross-Domain Recommendation}

As cross-domain recommendation systems evolved, a growing body of research examined the relationship between CDR and diversity. Initial studies primarily investigated whether CDR could enhance recommendation diversity. Several surveys~\cite{khan2017cross,cantador2015cross,fernandez2012cross} suggested that diversity could be a potential advantage of CDR systems. Building on these studies, Fernández-Tobías et al.~\cite{fernandez2016accuracy} conducted reproducible experiments to analyze the effect of CDR on diversity. Their results showed that existing CDR algorithms could produce more diverse recommendations depending on the target domain. More recent studies focused on developing diversity-aware approaches for CDR. For instance, CCDR~\cite{XLW22} constructed a diversified preference network to capture multiple facets of user interests. AD-CDMF~\cite{sun2022prick} introduced an adaptive diversity-regularized matrix factorization model. DivCDSR~\cite{chen2026divcdsr} explored diversity in a sequential scenario through domain homogeneity and information asymmetry. 

%% file: 7.conclusion.tex
\section{Conclusion}
This work investigated how user overlap shapes diversity in cross-domain recommendation. We identified two complementary diversity risks: concentrated exposure among overlapping users and excessive dominance of overlap-induced signals for non-overlapping users with sparse target-domain evidence. Based on this diagnosis, we proposed DivCDR, a model-agnostic Push--Pull framework. Pull constructs confidence-gated virtual positives to diversify overlap-supported exposure, while Push employs directional negative sampling to help non-overlapping users escape overlap-dense regions. Experiments across three datasets and five representative CDR backbones demonstrate consistent diversity improvements while largely preserving, and often improving, accuracy. 

%% file: acks.tex
\begin{acks}
This work was supported by NSFC/Research Grants Council of Hong Kong Joint Research Scheme (N\_HKBU214/24).
\end{acks}

%% file: 6.ethic.tex
\section*{Ethical Considerations}
DivCDR is evaluated on public benchmark datasets and uses interaction records rather than additional personal attributes. Cross-domain recommendation nevertheless relies on linking user behavior across domains, which may raise privacy and consent concerns when such linkage is not explicitly authorized. Diversity optimization can also change the balance between relevance and exposure; the reported offline gains should therefore not be interpreted as evidence of improved user satisfaction. Practical use should provide transparent data controls and monitor recommendation utility separately for overlapping and non-overlapping users.

%% file: paper.bib
@article{zhao2026sc,
  title={SC-VUG: Fairness-Aware Cross-Domain Recommendation for Non-Overlapping Users via Sparsity-Calibrated Virtual User Generation},
  author={Zhao, Yuhan and Chen, Weixin and Chen, Li and Pan, Weike},
  journal={ACM Transactions on Recommender Systems},
  year={2026},
  publisher={ACM, New York, NY}
}

@inproceedings{hu2026ids,
  title={From IDs to Semantics: A Generative Framework for Cross-domain Recommendation with Adaptive Semantic Tokenization},
  author={Hu, Peiyu and Lu, Wayne and Wang, Jia},
  booktitle={AAAI},
  volume={40},
  number={17},
  pages={14874--14882},
  year={2026}
}

@inproceedings{zhao2026double,
  title={The Double-Edged Sword of Knowledge Transfer: Diagnosing and Curing Fairness Pathologies in Cross-Domain Recommendation},
  author={Zhao, Yuhan and Chen, Weixin and Chen, Li and Pan, Weike},
  booktitle={WWW},
  pages={6865--6875},
  year={2026}
}

@inproceedings{chen2026divcdsr,
  title={DivCDSR: A Model-Agnostic Framework for Diverse Cross-Domain Sequential Recommendation},
  author={Chen, Shu and Zhao, Yuhan and Chen, Weixin and Pan, Weike and Chen, Li},
  booktitle={KDD},
  pages={437--448},
  year={2026}
}

@article{kaminskas2016diversity,
  title={Diversity, Serendipity, Novelty, and Coverage: a Survey and Empirical Analysis of Beyond-Accuracy Objectives in Recommender Systems},
  author={Kaminskas, Marius and Bridge, Derek},
  journal={TiiS},
  volume={7},
  number={1},
  pages={1--42},
  year={2016}
}

@article{wu2024result,
  title={Result Diversification in Search and Recommendation: A survey},
  author={Wu, Haolun and Zhang, Yansen and Ma, Chen and Lyu, Fuyuan and He, Bowei and Mitra, Bhaskar and Liu, Xue},
  journal={TKDE},
  volume={36},
  number={10},
  pages={5354--5373},
  year={2024}
}

@article{kunaver2017diversity,
  title={Diversity in Recommender Systems--A Survey},
  author={Kunaver, Matev{\v{z}} and Po{\v{z}}rl, Toma{\v{z}}},
  journal={KBS},
  volume={123},
  pages={154--162},
  year={2017}
}

@inproceedings{CCDR,
author = {Menglin, Kong and Wang, Jia and Pan, Yushan and Zhang, Haiyang and Hou, Muzhou},
title = {C²DR: Robust Cross-Domain Recommendation based on Causal Disentanglement},
year = {2024},
booktitle = {WSDM},
pages = {341–349}
}

@inproceedings{chen2025lasso,
  title={Lasso: Large Language Model-based User Simulator for Cross-Domain Recommendation},
  author={Chen, Yue and Yang, Susen and Zhang, Tong and Wang, Chao and Cheng, Mingyue and Lei, Chenyi and Li, Han},
  booktitle={RecSys},
  pages={207--216},
  year={2025}
}

@inproceedings{CDCDR,
author = {Li, Hanyu and Li, Jiayu and Ma, Weizhi and Sun, Peijie and Wu, Haiyang and Wang, Jingwen and Yang, Yuekui and Zhang, Min and Ma, Shaoping},
title = {CD-CDR: Conditional Diffusion-based Item Generation for Cross-Domain Recommendation},
year = {2025},
booktitle = {SIGIR},
pages = {1789–1798}
}

@article{khan2017cross,
  title={Cross Domain Recommender Systems: A Systematic Literature Review},
  author={Khan, Muhammad Murad and Ibrahim, Roliana and Ghani, Imran},
  journal={ACM Computing Surveys},
  volume={50},
  number={3},
  pages={1--34},
  year={2017},
  publisher={ACM New York, NY, USA}
}

@incollection{cantador2015cross,
  title={Cross-Domain Recommender Systems},
  author={Cantador, Iv{\'a}n and Fern{\'a}ndez-Tob{\'\i}as, Ignacio and Berkovsky, Shlomo and Cremonesi, Paolo},
  booktitle={Recommender Systems Handbook},
  pages={919--959},
  year={2015},
  publisher={Springer}
}

@inproceedings{fernandez2012cross,
  title={Cross-Domain Recommender Systems: A Survey of the State of the Art},
  author={Fern{\'a}ndez-Tob{\'\i}as, Ignacio and Cantador, Iv{\'a}n and Kaminskas, Marius and Ricci, Francesco},
  booktitle={Spanish Conference on Information Retrieval},
  volume={24},
  year={2012},
  organization={sn}
}

@article{sun2022prick,
  title={Prick the Filter Bubble: A Novel Cross Domain Recommendation Model with Adaptive Diversity Regularization},
  author={Sun, Jianshan and Song, Jian and Jiang, Yuanchun and Liu, Yezheng and Li, Jun},
  journal={Electronic Markets},
  volume={32},
  number={1},
  pages={101--121},
  year={2022},
  publisher={Springer}
}

@inproceedings{fernandez2016accuracy,
  title={Accuracy and Diversity in Cross-Domain Recommendations for Cold-Start Users with Positive-Only Feedback},
  author={Fern{\'a}ndez-Tob{\'\i}as, Ignacio and Tomeo, Paolo and Cantador, Iv{\'a}n and Di Noia, Tommaso and Di Sciascio, Eugenio},
  booktitle={RecSys},
  pages={119--122},
  year={2016}
}

@inproceedings{guo2024prompt,
  title={Prompt-Enhanced Federated Content Representation Learning for Cross-Domain Recommendation},
  author={Guo, Lei and Lu, Ziang and Yu, Junliang and Nguyen, Quoc Viet Hung and Yin, Hongzhi},
  booktitle={WWW},
  pages={3139--3149},
  year={2024}
}

@inproceedings{liu2024graph,
  title={Graph Disentangled Contrastive Learning with Personalized Transfer for Cross-Domain Recommendation},
  author={Liu, Jing and Sun, Lele and Nie, Weizhi and Jing, Peiguang and Su, Yuting},
  booktitle={AAAI},
  volume={38},
  number={8},
  pages={8769--8777},
  year={2024}
}

@article{tian2024privacy,
  title={Privacy-Preserving Cross-domain Recommendation with Federated Graph Learning},
  author={Tian, Changxin and Xie, Yuexiang and Chen, Xu and Li, Yaliang and Zhao, Xin},
  journal={TOIS},
  volume={42},
  number={5},
  pages={1--29},
  year={2024},
  publisher={ACM New York, NY}
}

@inproceedings{chen2025leave,
  title={Leave No One Behind: Fairness-Aware Cross-Domain Recommender Systems for Non-Overlapping Users},
  author={Chen, Weixin and Zhao, Yuhan and Chen, Li and Pan, Weike},
  booktitle={RecSys},
  pages={226--236},
  year={2025}
}

@inproceedings{GLC13,
  title={Cross-Domain Recommendation via Cluster-Level Latent Factor Model},
  author={Gao, Sheng and Luo, Hao and Chen, Da and Li, Shantao and Gallinari, Patrick and Guo, Jun},
  booktitle={PKDD},
  pages={161--176},
  year={2013}
}

@article{ZCL,
  title={Denoising and Augmented Negative Sampling for Collaborative Filtering},
  author={Zhao, Yuhan and Chen, Rui and Lai, Riwei and Han, Qilong and Song, Hongtao and Chen, Li},
  journal={TORS},
  volume={3},
  number={4},
  pages={1--23},
  year={2025},
  publisher={ACM, New York, NY}
}

@inproceedings{XLW22,
  title={Contrastive Cross-Domain Recommendation in Matching},
  author={Xie, Ruobing and Liu, Qi and Wang, Liangdong and Liu, Shukai and Zhang, Bo and Lin, Leyu},
  booktitle={KDD},
  pages={4226--4236},
  year={2022}
}

@article{ZWC21,
  title={Cross-Domain Recommendation: Challenges, Progress, and Prospects},
  author={Zhu, Feng and Wang, Yan and Chen, Chaochao and Zhou, Jun and Li, Longfei and Liu, Guanfeng},
  journal={arXiv preprint arXiv:2103.01696},
  year={2021}
}

@inproceedings{LCL24,
  title={User Distribution Mapping Modelling with Collaborative Filtering for Cross Domain Recommendation},
  author={Liu, Weiming and Chen, Chaochao and Liao, Xinting and Hu, Mengling and Su, Jiajie and Tan, Yanchao and Wang, Fan},
  booktitle={WWW},
  pages={334--343},
  year={2024}
}

@inproceedings{SG08,
  title={Relational Learning via Collective Matrix Factorization},
  author={Singh, Ajit P and Gordon, Geoffrey J},
  booktitle={KDD},
  pages={650--658},
  year={2008}
}

@article{WYW24,
  title={Making Non-Overlapping Matters: An Unsupervised Alignment Enhanced Cross-Domain Cold-Start Recommendation},
  author={Wang, Zihan and Yang, Yonghui and Wu, Le and Hong, Richang and Wang, Meng},
  journal={TKDE},
pages={2001--2014},
volume={37},
  year={2025},
}

@article{WYZ25,
author = {Wang, Hao and Yin, Mingjia and Zhang, Luankang and Zhao, Sirui and Chen, Enhong},
title = {MF-GSLAE: A Multi-Factor User Representation Pre-Training Framework for Dual-Target Cross-Domain Recommendation},
year = {2025},
volume = {43},
number = {2},
journal = {TOIS}
}

@inproceedings{LLL20,
  title={Cross Domain Recommendation via Bi-Directional Transfer Graph Collaborative Filtering Networks},
  author={Liu, Meng and Li, Jianjun and Li, Guohui and Pan, Peng},
  booktitle={CIKM},
  pages={885--894},
  year={2020}
}

@inproceedings{MSJ17,
  title={Cross-Domain Recommendation: An Embedding and Mapping Approach},
  author={Man, Tong and Shen, Huawei and Jin, Xiaolong and Cheng, Xueqi},
  booktitle={IJCAI},
  volume={17},
  pages={2464--2470},
  year={2017}
}

@inproceedings{ZHP22,
  title={RecBole 2.0: Towards a More Up-to-Date Recommendation Library},
  author={Zhao, Wayne Xin and Hou, Yupeng and Pan, Xingyu and Yang, Chen and Zhang, Zeyu and Lin, Zihan and Zhang, Jingsen and Bian, Shuqing and Tang, Jiakai and Sun, Wenqi and others},
  booktitle={CIKM},
  pages={4722--4726},
  year={2022}
}

@inproceedings{GDH22,
  title={Self-Guided Learning to Denoise for Robust Recommendation},
  author={Gao, Yunjun and Du, Yuntao and Hu, Yujia and Chen, Lu and Zhu, Xinjun and Fang, Ziquan and Zheng, Baihua},
  booktitle={SIGIR},
  pages={1412--1422},
  year={2022}
}

@inproceedings{ZCL23,
author = {Zhao, Yuhan and Chen, Rui and Lai, Riwei and Han, Qilong and Song, Hongtao and Chen, Li},
title = {Augmented Negative Sampling for Collaborative Filtering},
booktitle = {RecSys},
pages = {256–266},
year = {2023}
}

@inproceedings{HYC17,
  author       = {Cheng{-}Kang Hsieh and
                  Longqi Yang and
                  Yin Cui and
                  Tsung{-}Yi Lin and
                  Serge J. Belongie and
                  Deborah Estrin},
  title        = {Collaborative Metric Learning},
  booktitle    = {WWW},
  pages        = {193--201},
  year         = {2017}
}

@article{DQY20,
  title={Simplify and Robustify Negative Sampling for Implicit Collaborative Filtering},
  author={Ding, Jingtao and Quan, Yuhan and Yao, Quanming and Li, Yong and Jin, Depeng},
  journal={NeurIPS},
  volume={33},
  year={2020}
}

@inproceedings{RFG12,
  author       = {Rendle, Steffen and Freudenthaler, Christoph and Gantner, Zeno and Schmidt-Thieme, Lars},
  title        = {{BPR:} Bayesian Personalized Ranking from Implicit Feedback},
  booktitle    = {UAI},
  pages        = {452--461},
  year         = {2009}

}

@inproceedings{ZZY22,
  title={Learning Explicit User Interest Boundary for Recommendation},
  author={Zhuo, Jianhuan and Zhu, Qiannan and Yue, Yinliang and Zhao, Yuhong},
  booktitle={WWW},
  pages={193--202},
  year={2022}
}

@inproceedings{YYX22,
  title={Are Graph Augmentations Necessary? Simple Graph Contrastive Learning for Recommendation},
  author={Yu, Junliang and Yin, Hongzhi and Xia, Xin and Chen, Tong and Cui, Lizhen and Nguyen, Quoc Viet Hung},
  booktitle={SIGIR},
  pages={1294--1303},
  year={2022}
}

@inproceedings{ZMH21,
  title={RecBole: Towards a Unified, Comprehensive and Efficient Framework for Recommendation Algorithms},
  author={Zhao, Wayne Xin and Mu, Shanlei and Hou, Yupeng and Lin, Zihan and Chen, Yushuo and Pan, Xingyu and Li, Kaiyuan and Lu, Yujie and Wang, Hui and Tian, Changxin and others},
  booktitle={CIKM},
  pages={4653--4664},
  year={2021}
}

@article{ZLZY23,
  author       = {Tianzi Zang and
                  Yanmin Zhu and
                  Haobing Liu and
                  Ruohan Zhang and
                  Jiadi Yu},
  title        = {A Survey on Cross-domain Recommendation: Taxonomies, Methods, and Future Directions},
  journal      = {TOIS},
  volume       = {41},
  number       = {2},
  pages        = {42:1--42:39},
  year         = {2023},
}

@inproceedings{zhu2019dtcdr,
  title={DTCDR: A Framework for Dual-Target Cross-Domain Recommendation},
  author={Zhu, Feng and Chen, Chaochao and Wang, Yan and Liu, Guanfeng and Zheng, Xiaolin},
  booktitle={CIKM},
  pages={1533--1542},
  year={2019}
}

@article{yuan2022tenrec,
  title={Tenrec: A Large-Scale Multipurpose Benchmark Dataset for Recommender Systems},
  author={Yuan, Guanghu and Yuan, Fajie and Li, Yudong and Kong, Beibei and Li, Shujie and Chen, Lei and Yang, Min and Yu, Chenyun and Hu, Bo and Li, Zang and others},
  journal={NeurIPS},
  volume={35},
  pages={11480--11493},
  year={2022}
}
